# EVOLVABLE AUTONOMIC MANAGEMENT

R. KAMAL[1]

**Abstract.** Autonomic management is aimed at adapting to uncertainty. Hence, it is devised as m-connected k-dominating set problem, resembled by dominator and dominate, such that dominators are resilient up to m-1 uncertainty among them and dominate are resilient up to k-1 uncertainty on their way to dominators. Therefore, an evolutionary algorithm GENESIS is proposed, which resolves uncertainty by evolving population of solutions, while considering uncertain constraints as sub-problems, started by initial populations by a greedy algorithm AVIDO. Theoretical analysis first justifies original problem as NP-hard problem. Eventually, the absence of polynomial time approximation scheme necessitates justification of original problem as multiobjective optimization problem. Furthermore, approximation to Pareto front is verified to be decomposed into scalar optimization sub-problems, which lays out the theoretical foundation for decomposition based evolutionary solution. Finally, case-study, feasibility analysis and exemplary implication are presented for evolvable autonomic management in combined cancer treatment with in-vivo sensor networks. Keywords: Autonomic Management, Connected Dominating Set, Evolutionary Computation.

AMS Subject Classification: The author(s) should provide AMS Subject Classification numbers using the link http://www.ams.org/mathscinet/msc/msc2010.html

## 1. Introduction

**1.1. Motivation.** The penetration of Smart-devices and the explosive growth of internet have resulted in complex multi-vendor environment, where manufacturers, operators, and service providers seem to monetize by deploying emerging technologies on this situation[15][14][13]. Accordingly, the increasing demand of personalized services and conjugal business models have striven for more advanced and complex management issues[15][14][13]. In this context, conventional network and service management paradigms, seems to be merely rigid to not only specific network topologies, bult also manual deployment and management, which might often incorporate inflexibility, high labor and time constraints to operational cost[2].

In this context, autonomic paradigm (i.e computing, management) has proliferated as an enabling technology to attenuate human intervention by a self-governing behavior, which is merely within the constraints of overall objective, manager as a whole seeks to achieve[16]. Hence, an autonomic control loop is maintained, so that system, (a) senses uncertainty in managed element and environment and (b) analyzes information to ensure objectives are met, (c) hastens resilience, if objectives are not met, (d) finally, observes the result[2]. Accordingly, an autonomic approach simplifies management by facilitating resilience in uncertainties to automate overall decision making for optimizing operation.

**1.2. Autonomic Management Problem.** Hence, *Autonomic Management problem* (Section II)(Definition 6) is formulated by aiming resilience to uncertainty through $m$-connected $k$-dominating set with multiple uncertainty constraints, such that, (a) each dominate is resilient up to $k$ uncertainties on its way to dominators, (b) each dominator is resilient up to $m$-1 uncertainties on it way to dominators.

[1] Department of Computer Engineering, Kyung Hee University, South Korea,
e-mail: ROSSI.KAMAL.R@ieee.org

1.3. **Proposed Evolutionary Scheme.** In this context, an evolutionary scheme (Section III) is proposed, that facilitates resilience to uncertainty by evolving solutions of multiple uncertainty metrics. In this process, first, greedy algorithm *AVIDO* constructs $m$-connected $k$-dominating sets with an uncertainty metric. Then, multi-objective evolutionary algorithm *GENESIS* generates Pareto optimal solutions, by evolving population of solution of each subproblem, where each subproblem corresponds to an uncertainty metric. Hence, solutions for different uncertainty metrics (generated by *AVIDO*) construct the initial population of *GENESIS*.

1.4. **Theoretical Analysis.** Accordingly, theoretical analysis (Section IV) justifies the following-

(a)*Autonomic Management* is as a NP-hard problem. Eventually, the absence of polynomial time approximation scheme necessitates justification of this problem as multiobjective optimization problem (MOP).

(b)Approximation to Pareto front (i.e MOP solutions, which are superior to other solutions in the search space for all objectives) is decomposable into scalar optimization subproblems, which lays out the theoretical foundation of decomposition based evolutionary solution.

(c)AVIDO outperforms state-of-the art dominating set research with the cost analysis and proof of outcome of different rounds. It is followed by comparative summarization of *GENESIS* and *MOE-A/D*.

1.5. **Empirical Results.** Finally, empirical results (Section V) can be summarized as follows-

(a)Rigorous simulation on Sinalgo first verifies the supremacy of AVIDO in terms of 'impact of transmission range, network size on backbone size', 'impact of cost constraint on backbone size and maximum length and impact of fault-tolerance on backbone size'. It is followed by substantiation of *GENESIS* over *AVIDO* in terms of uncertainty metrics, namely presence and absence of transmission rate, respectively.

(b)Extensive case-study is performed to analyze effectiveness of combined cancer treatment with radiotherapy and chemotherapy, with the influence of hyperthermia. In this context, the feasibility of the implication of in-vivo sensor networks on combined cancer treatment is analyzed with special consideration to in-vivo hotspot creation and uncertain communication. Finally, an exemplary scenario illustrates the application of evolvable autonomic management in combined cancer treatment under uncertain in-vivo environment.

1.6. **Organization.** The paper is organized as follows- (a) *Autonomic Management* problem is formulated on Section II, (b) Proposed evolutionary scheme is presented on Section III, (c) Theoretical analysis is presented on Section IV, (d) Empirical results are presented on Section V, followed by conclusion on Section VI.

## 2. Problem Formulation

In this section, *Autonomic Management* problem is formulated, preceded and followed by preliminary definitions and integer linear programming formulation of $m$-connected $k$-dominating sets in Unit Disk Graph(UDG) model.Consequently, basic notations, uncertainty and resilence factors of UDG models are summarized in Table 1, 2, 3, respectively, which assist in problem formulation and thereafter.

**Definition 1.** *Unit Disk Graph (UDG):A unit disk is a disk having diameter one. A unit disk graph (UDG) is a set of unit disks in Euclidian plane. Each node is in the center of a unit disk. An edge E(u,v) exists between two nodes u and v, if disks associated with u and v intersects each other.*

**Definition 2.** *$D(G)$ is a dominating set of G if $\forall v \in G$, either $v \in D(G)$ or $\exists u$ such that $(u,v) \in E(G)$.*

Table 1. Notations in UDG

| Notation | Description |
|---|---|
| G | A graph |
| V(G) | Vertices of a graph |
| (u,v) | A pair of vertices of a graph |
| E(G) | An edge between pair of vertices in a graph G |
| I | Independent Set |
| M | Maximal Independent Set |
| D | Dominating Set |
| UDG | Unit Disk Graph |

Table 2. Uncertainty factors in UDG

| Definition | Description |
|---|---|
| Bad point | A vertex v∈V(G) is a bad point, if subgraph induced by G-v is not 2-connected. |
| Cut vertex | A vertex v∈V(G) is a cut vertex if graph G¥ v is disconnected. |
| Leaf block | A leaf block of a connected graph G is a subgraph such that it is a block and contains one cut-vertex of G |
| Independent set | I(G)⊆V(G) is an independent set of G if ∀(u,v), no edge exists between u and v. |
| Maximal Independent set | An independent set M is a maximal independent set if no v∈(G ¥M) can be added to M. If any v∈(G-M) is added, it is not an independent set anymore. |

**Definition 3.** *D(G) is a connected dominating set of G if (a) D(G)⊆V(G) is a dominating set of G and (b) graph induced by D(G) is connected.*

**Definition 4.** *A dominating set D(G)⊆V(G) is k-dominating if ∀v∈V(G) ¥ D(G), v is adjacent to at least k nodes in D(G).*

**Definition 5.** *A dominating set D(G)⊆V(G) is m-connected if graph induced by D is m-connected. It means that D is connected after $m-1$ dominators are removed.*

**Definition 6.** *Autonomic Management: Given a unit disk graph $G(V,E)$, two positive integers m, k, and uncertainty (Table 2) and resilience (Table 3) factors,* Autonomic Management *problem is aimed at finding a subset D ⊆V(G) such that (a)each vertex v ∈(V(G)¥D) is k-dominated by at least one vertex in D such that guaranteed routing is maintained (b)D is m-connected*

Table 3. Resilence factors in UDG

| Definition | Description |
|---|---|
| Good point | A vertex v∈V(G) is a good point, if subgraph induced by G¥v is still 2-connected. |
| Maximal connected subgraph | A block is a maximal connected subgraph of G that does not have any cut-vertex. |

*Hence,* Autonomic Management *is aimed at facilitating m-connected k-dominating set with minimum constraints satisfying (a) and (b).*

2.1. ILP formulation. Now, ILP is formulated for *Autonomic Management* problem. In this context, at first, ILP is formulated for connected dominating set(1)-(11). This portion of ILP formulation is inspired from CDS construction by spanning tree. It is followed by ILP formulation for *m*-connected *k*-dominating set with minimum cost constraint(12)-(15).

$$\min c \tag{1}$$

subject to

$$\sum_{i \epsilon V} a_i \leq D \tag{2}$$

$$b_{ij} \leq na_i i\epsilon V \ \text{¥} \{1\} \tag{3}$$

$$\sum_{j \epsilon V} b_{ij} < 1 + (n-1)a_1 \tag{4}$$

$$\sum_{j \epsilon V} b_{ij} \geq 1 \tag{5}$$

$$\sum_{i,j \epsilon V; i=j} b_{ij} = n - 1 \tag{6}$$

$$u_i = 1 \tag{7}$$

$$u_j \leq ni\epsilon V \ \text{¥} 1 \tag{8}$$

$$u_j \geq 2i\epsilon V \ \text{¥} 1 \tag{9}$$

$$u_i - u_j + 1 \leq (n-1)(1 - b_{ij}), i, j \epsilon V \ \text{¥} 1 \tag{10}$$

$$a_i, b_{ij} \epsilon \{0, 1\} i, j \epsilon V, i = j \tag{11}$$

$$\min c \tag{12}$$

subject to

$$\sum_{t \epsilon N(s)} x_t \geq mx_s, \forall s \epsilon V \tag{13}$$

---

**Algorithm 1 AVIDO**

1. Round 1: MIS Construction
2. INPUT: Color all nodes as WHITE node
3. Choose a node with maximum cardinality and select as root of MIS and color it as BLACK.
4. Color the neighbors of MIS node as GREY node
5. while There is no WHITE node do
6.     Choose the WHITE node that has the most grey neighbors and color it BLACK as MIS node
7.     Color the neighbors of new created black node as GREY
8. end while
9. OUTPUT: BLACK MIS nodes and other GREY nodes
10. Round 2: 1-connected 1-dominating set construction
11. INPUT: initially D is empty and BLACK MIS nodes, GREY nodes are present
12. for Every pair of BLACK nodes u and v with d(u,v)<=4 do
13.     Compute shortest path p(u,v) and color all intermediate GREY nodes of p(u,v) as BLACK
14.     Add u, v and intermediate nodes to D
15. end for
16. OUTPUT: D contains all BLACK nodes (MIS and connected nodes)
17. Round 3:1-connected k-dominating set
18. INPUT: 1-connected 1-dominating set
19. Remove MIS from the graph,G= G-$M_1$
20. for i=2 to k do
21.     Construct $M_i$ in G-$M_1 \cup M_2 \cup ...M_i - 1$
22.     D= D $\cup$ $M_i$ (Following Round 2)
23. end for
24. OUTPUT:1-connected k-dominating set
25. Round 4:2-connected k-dominating set
26. INPUT:1-connected k-dominating set
27. Find all blocks in 1-connected k-dominating set
28. while D is not 2-connected do
29.     Compute all blocks in graph
30.     Add all intermediate nodes of shortest path that (a)connects leaf block in D to other part of D (b)does not have any nodes in D except two endpoints
31. end while
32. OUTPUT:2-connected k-dominating set
33. Round 5: 3-connected k-dominating set
34. INPUT:2-connected k-dominating set
35. while There is no badpoint do
36.     Convert bad point to good point by moving from G-D to D, such that no new bad point is created
37. end while
38. OUTPUT:3-connected k-dominating set

$$\sum_{t \epsilon N(s)} x_t \geq k(1 - x_s), \forall s \epsilon V \tag{14}$$

$$x_s \epsilon \{0, 1\}, \forall s \epsilon V \tag{15}$$

Let, $a_i$, $i \in V$ be a binary decision variable indicating whether $i$ beolongs to CDS. Let, $b_{ij}$, $i,j \in V$ and i=j, be a binary decision variable indicating whether that edge is connected to CDS. Let, $c$ represents the cost of CDS.

The objective(1) is to minimize the cost. Constraints are numbered from (2) to (11).(1) means that size of CDS is less than or equal to D.(2) indicates that only the vertices in connected dominating set have outgoing edges.(3)indicates that first node can have edge going out even if it is a leaf node not in CDS. (4)means that first node should have at least one edge going out even if it is a leat.(5)indicates that the spanning tree must have (n-1)edges. (6-11)are used to avoid cycles with inspiration from classical MTZ(Miller,Tucker and Zemlin) formulation.

Let,$x_s$ be a binary variable. $x_s$=1,if s is a dominator. Otherwise, $x_s$=0, if s is a dominate.

The objective (12) is to minimize the cost. Constraints are numbered from (13) to (15). The first restriction is, that there exists m disjoint paths between any pair of dominators. Constraint(13) indicates the first restriction. It shows that, if a certain vertex s is set to dominator ($x_s$=1), then vertex s is adjacent to at least m different dominators and thereby form m-connected CDS. The second restriction is that if a vertex is a dominate, then it has at least k dominators. Constraint(14) indicates the second restriction. It shows that if a certain vertex s is a dominate ($x_s$=0), then there exists at least k adjacent dominators to the vertex s. Moreover, constraint(15) indicates that any node s can be either dominator or dominate, so decision variable $x_s$ is either 1 or 0, respectively.

## 3. Proposed Evolutionary Mechanism

In this section, an evolutionary scheme is proposed, that facilitates evolvable autonomic management by enabling resilience to uncertainty. In this process, an evolutionary algorithm *GENESIS* (Algorithm 2) constructs m-connected k-dominating sets with multiple uncertainty constraints by evolving initial solutions for different uncertainty constraints, generated by a greedy algorithm *AVIDO* (Algorithm 1).

3.1. **Greedy Algorithm** *AVIDO*. At first, *AVIDO* (Algorithm 1) is described, which constructs m(<=3)-connected k(<=3)-dominating sets for UDG with a minimum uncertainty constraint. It is a round based algorithm. At first, MIS is constructed in unit disk graph model(Round 1). Then, nodes are connected to MIS nodes to construct 1-connected 1-dominating set considering uncertainty concept [9](Round 2). Then, subsequent $k-1$ MISs are added to CDS to construct 1-connected $k$-dominating set (Round 3)(Fig. 2(b)). 1-connected k-dominating set is iteratively augmented to construct 2-connected k-dominating set [24](Round 4)(Fig.2(c)). At last, 2-connected CDS is turned to 3-connected by turning all bad points to good points using the strategy of [17](Round 5)(Fig.2(d)). As a result, AVIDO produces m(<=3)-connected k(<=3)-dominating sets.

3.2. **Multi-Objective Evolutionary Algorithm** *GENESIS*. Now, evolutionary algorithm *GENESIS* (Algorithm 2)(Fig. 1) is proposed, which constructs Pareto front by evolving initial solutions by *AVIDO*. At first, Pareto front is assumed to empty (step 2). Now, the relation among different uncertainty metrics is calculated (step 3). Let, L be number of uncertainty metrics. Each sub-problem is related to an uncertainty metric. Then, maximum T relevant uncertainty metrics of each uncertainty metric is calculated. Now, initial population from the solution of each uncertainty metric is calculated (step 4) from AVIDO. Then, each sub-problem/uncertainty metric is processed (step 5). Then, a new solution by crossover/mutation operation of solutions of two chosen metrics, who are assumed to be relevant(step 6). Improvement/repair is performed on new solution, if possible(step 7).

For example, repair is performed as follows, for two uncertainty metrics a and b, $\frac{\sum_{i=1}^{T} \lambda_i \cdot a_i^2}{b^2}$

When a solution with respect to uncertainty metric a is desired, it is to be maximized. Meanwhile, the relevant uncertainty metric b should be as less as possible.

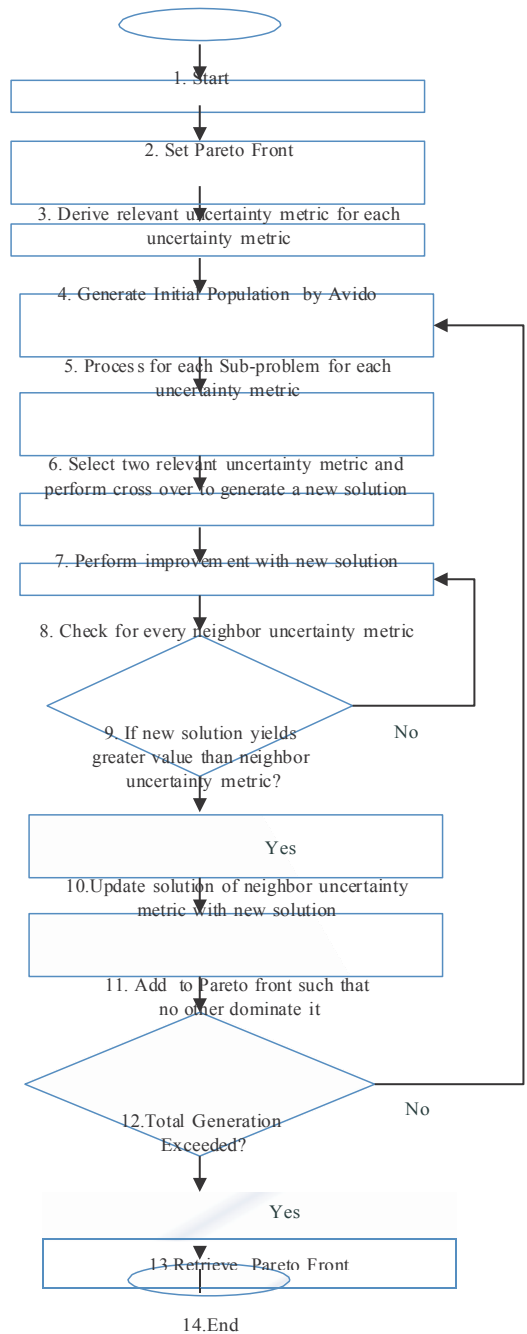

Figure 1. Flowchart of Genesis Algorithm

The repair is performed for each relevant sub-problem (step 8). If any relevant sub-problem is found, whose solution is less than new solution(step 9), then that solution is given the value of new solution(step 10). Then, it is added to to Pareto front(step 11). Then, it is checked, whether total generation is reached (step 12). Otherwise, step 5 is executed again to deal with N sub-problems. At last, Pareto front contains the output solutions or output population(step 13).

As mentioned before, *GENESIS* is an instance of MOEA/D [27].

Table 4. Comparison of *AVIDO* problem with proposals [17],[18] and [9]

| Proposal | Fault-tolerance | Minimum Routing | 3D space Integration | Approximation Bound |
|---|---|---|---|---|
| [17] | Present | Absent | Absent | 520/3 |
| [18] | Absent | Absent | Present | 14.937 |
| [9] | Absent | Present | Absent | No PTAS exists. |
| *AVIDO* | Present | Present | Absent | No PTAS exists. |

## 4. Theoretical Analysis

In this section, *Autonomic Management* is justified as NP-hard and multi-objective optimization problem, followed by substantiation of the amenability of Pareto front (i.e MOP optimal solution) to decomposition and evolutionary scheme. Moreover, the supremacy of *AVIDO* and *GENESIS* over state-of-the art are also summarized in Table 4 and 5,resppectively in addition to cost analysis and outcome of different rounds of *AVIDO*.

*Lemma* 1. *Autonomic Management is a NP-hard Problem*

*Proof.* CDS-construction is a NP-hard problem in UDG[8]. Hence, $m$-connected $k$-dominating set with uncertainty constraints,and thereby *Autonomic Management* is a NP-hard problem.

*Lemma* 2. *No PTAS exists for* Autonomic Management *problem*

*Proof.* There is no PTAS available for weighted CDS-construction[8]. It yields that no PTAS exists for $m$-connected $k$-dominating set with uncertainty contraint. Hence, no PTAS exists for *Autonomic Management*.

*Lemma* 3. MIS M is created after round 1 of AVIDO.

*Proof.* In round 1, when a node joins M, its neighbors are colored gray. Next, unexplored white node joins M. So, there is no possibility of gray node to join MIS. So, no two neighbors are included in M. Also, round 1 ends when there is no white node. So, there cannot be any node left to be added to MIS M. So, MIS M is created after round 1 of *AVIDO*.

*Lemma* 4. CDS D is created after round 2 of AVIDO.

*Proof.* At round 2, all intermediate grey nodes in the path of two MIS nodes(where ROUTING-COST is below THRESHOLD) are colored black. Those black nodes (intermediate nodes and MIS nodes) construct connected dominating set.

*Lemma* 5. 1-connected k-dominating set is created after round 3 of *AVIDO*

*Proof.* Let, G, D, I be connected graph, connected dominating set and maximal independent set respectively. After MIS and then CDS construction in first two rounds, k-1 subsequent MIS are added to CDS in third round. As a result of it, G-D nodes are k dominated by D nodes. That means each node of G-D is connected to k nodes of D. So, 1-connected k-dominating set is found.

*Lemma* 6. 2-connected dominating is created after round 4 of *AVIDO*. *Proof:* At the end of round 4, all dominator nodes are in same block, so that dominators are 2-connected[24]. So, we get 2-connected k-dominating set at the end of round 4.

*Lemma* 7. If $v \epsilon G$ is a good point, subgraph G- {v} is 2-connected[24].

*Lemma* 8. A 2-connected graph without any bad point is 3-connected.

*Proof.* A graph G is 3-connected if we need to remove at least three nodes to disconnect G. For example, v be a good point in 2-connected graph G'. From Lemma 11, G'-(v) is still 2-connected. So, we need to remove at least 2 nodes to disconnect G'. So, we can say G' is 3-connected.

*Lemma* 9. 3-connected dominating is created after round 5 of *AVIDO*

*Proof.* In round 5, all bad points are converted to only good points by transferring some nodes from non-dominator set to dominator set. At the end, 3-connected 3-dominating set is created [17].

*Lemma* 10. Approximation bound for uncertainty constraint in *AVIDO* is $d(u, v) \leq 4$.

*Proof.* We have considered uncertainty as a constraint in *AVIDO*.

Let, $u, v$ are two nodes to be connected to CDS with minimum uncertainty and $d(u, v)$ is distance between them. Threfore, $d(u, v) \leq 4$ is obtained by following[9] and by considering the following facts.

Let, $a, b$ are two MIS nodes such that $d(a, b) = 2$, it means that no MIS nodes are more than two hops from each other. Now, $u, v$ are two nodes such that $d(u, v) \leq 4$. Let us connect $u, v$ to connected dominating set. Therefore, $u$ can be one hop away from $a$ and $v$ can be one hop away from $b$.

Therefore, $d_D(u, v) <= d(a, b) + 2$
$d_D(u, v) \leq 2 + 2 \leq 4$
$d_D$ represents the distance between $u, v$ when they are connected in *CDS*.

*Lemma* 11. *Autonomic Management* is a Multi-objective Optimization [27] Problem (MOP).

*Proof. Autonomic Management* involves simultaneous optimization of incommensurable and often correlated objectives for different uncertainty constraints. Considering $x$ decision vector, $X$ parameter space, $y$ objective vector and $Y$ is objective space, *Autonomic Management* maps a tuple of $m$ decision variables to a tuple of $n$ objectives as follows

$$min/max \ y = f(x) = (f_1(x), f_2(x), \ldots, f_n(x)) (1) \qquad (16)$$

$$subject to \ x = (x_1, x_2, \ldots, x_n) \epsilon X \ (2) \qquad (17)$$

$$y = (y_1, y_2, \ldots, y_n) \epsilon Y \ (3) \qquad (18)$$

Hence, *Autonomic Management* is turned out to be a Multi-objective optimization problem.

*Lemma* 12. MOP *Autonomic Management* demands a Pareto optimal solution.

*Proof.* Likewise general MOP, *Autonomic Management* seeks a set of alternative solutions, rather than a single optimal solution. These solutions need to be superior to other solutions in the search space, by considering all objectives[27].

Given a maximization problem with two decision vectors for two uncertainty metrics, for example, $x_1, x_2,$ and a is said to dominate b iff

$$\forall i \epsilon \{1, .., n\} : f_i(x_1) \geq f_i(x_2) \land \forall \epsilon j \{1, .., n\} : f_j(a) \geq f_j(x_2) (4) \qquad (19)$$

Hence, non-dominated decision vectors, who are not nominated by other decision vector of a given set with the entire search space, constitute the optimal solution. Hence, Pareto optimal front is the desired solution of *Autonomic Management* problem.

*Lemma* 13. *Autonomic Management* is aminable to MOEA/D

*Proof.* MOEA/D decomposes MOP into $N$ scalar optimization subproblems and solves the subproblems by evolving a population of solutions. Each generation maintains a population of best solutions of each sub-problem. Hence, each sub-problem (with any uncertainty metric) of *Autonomic Management* problem is optimized by information from neighboring subproblems with corresponding uncertainty metrics. Hence, *Autonomic Management* is amenable to MOEA/D.

*Lemma* 14. Weighted sum-based decomposition approximates Pareto front of MOP-based *Autonomic Management* problem.

*Proof.* Weighted sum approach considers combinaton of the different objectives for uncertainty metrics. Let, $\lambda^1,....\lambda^N$ be weight vectors relevant to subproblems of different uncertainty metrics. Therefore, optimization problem is decomposed into $N$ scalar optimization subproblems by using this approach. The objective function of the $j$-th subproblem is

$$max g^{ws}(x|\lambda^j) = \sum_{i=1}^{n} \lambda_i^j f_i(x) \qquad (20)$$

It represents that if $\lambda^i$ and $\lambda^j$ are two uncertainty metrics closer to each other, optimal solution of their sub-problems $g^{ws}(x|\lambda^i)$ and $g^{ws}(x|\lambda^j)$ are closer to each other. So, one metric, having weight vector closer $i$, should be useful in optimizing other metric's solution, $g^{ws}(x|\lambda^i)$.

Hence, weighted sum-based decomposition approximates pareto front of MOP *Autonomic Management* problem.

## 5. Empirical Results

In this section, the performance of *AVIDO* and *GENESIS* is evaluated by conducting simulation on Sinalgo[12] simulator. Sinalgo is chosen, as it supports UDG network models used in our proposal. It is followed by an implication of evolvable autonomic management approach on combined cancer treatment with the advent of in-vivo body sensor networks.

### 5.1. Performance evaluation of AVIDO.

5.1.1. *Simulation Settings.* In the simulation, nodes are randomly deployed in a 100×100 plane. The number of nodes ranges from 50 to 150. One hundred connected UDGs are randomly generated in this simulation setup. All nodes are assumed to have same transmission range. A random value between 0.0 and 0.8 is assigned as the transmission rate between the nodes.

5.1.2. *Impact of Transmission Range and Network Size on Backbone Size.* Fig. 2(a) shows how the backbone size changes with the transmission range and network size. As the transmission range increases, the CDS size decreases because CDS nodes can dominate more non-CDS nodes and fewer nodes are needed to construct the CDS. As the network size increases, the CDS size increases as a larger CDS is needed to dominate the non-CDS nodes.

Table 5. Comparison of GENESIS with MOEA/D

| Issue | MOEA/D | MOE-Autonomic-D-WS |
|---|---|---|
| MOP | $Eq(1)$ | In-vivo routing with multiple constraints |
| Scalar Optimization Technique | Tchebycheff Approach | Weighted Sum Approach |
| Objective Function | $g^{te}(x\|\lambda^j, z*) = max_{1\leq i\leq m}\{\lambda_i^j\|f_i(x) - z_{i*}\}$ | $maxg^{ws}(x\|\lambda^j) = \sum_{i=1}^{m} \lambda_i^j f_i(x)$ |
| Neighbor Updating | Using new, smaller solution | Using new, greater solution |
| Genetic Operator | Random | Two Relevant routing contraints |
| Terminal Condition | Not defined | Generation above threshold |
| Improvement Condition | Not defined | $[\frac{\sum_{i=1}^{i=m}\lambda_i.R_E}{D^2}]^2$ |
| Application | Knapsack problem. | In-vivo routing efficiency. |

5.1.3. *Impact of uncertainty cost constraint both backbone size and maximum routing length.* Fig.2(b) shows the impact of uncertainty constraint on backbone size. When network size increases, 'With uncertainty cost constraint' generates larger backbone than 'without uncertainty cost constraint'. Because, it needs more nodes to add to CDS to generate shortest path in CDS for node pairs' outside CDS.

Fig. 2(c) shows the impact of uncertainty constraints on the maximum cost. 'With uncertainty cost' generates less 'maximum cost' than 'without uncertainty cost'. When using 'with uncertaintly cost', a node has a high probability of connecting to more neighbors, which does not increase uncertainty cost. Therefore, uncertainty cost constraint increases the backbone size, but decreases the maximum uncertainty cost for every node.

5.1.4. *Impact of fault-tolerance on backbone size.* Fig. 2(d) shows how backbone size changes with resilence. When the backbone has no resilience (1-connected, 1-dominating set), the backbone size is at its minimum. To construct 1-connected, 2-dominating set, one more MIS need to be added to the CDS. As such, the CDS size is also increased with improvement in the resilience. To construct a 2-connected 2-dominating set, it is necessary to augment the backbone by adding nodes to connect the leaf block in the backbone to other block(s). As a result, the backbone size increases. To construct 3-connected 3-dominating set, the backbone size also increases. There are two reasons for this. Firstly, non-CDS nodes have to move to the CDS to convert bad points to good points. Secondly, more MIS nodes have to be added to the CDS.

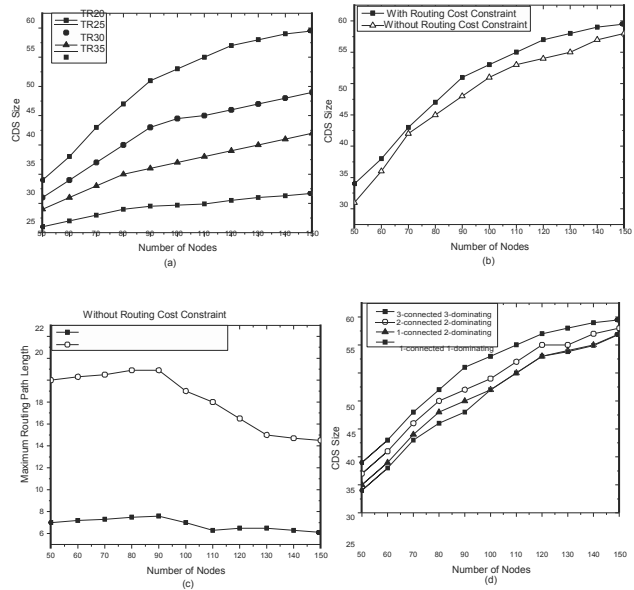

Figure 2. (a)Changes of CDS size with increase of number of nodes for different transmission ranges. Impact of uncertainty cost constraint on (b)CDS size and (c) maximum cost (d) Impact of fault-tolerance on backbone size

Table 6. Evolutionary Computation Parameters used in Simulation

| Evolutionary Computation Parameter | Value |
|---|---|
| Population size/total number of subproblems | 2 |
| Total generation | 50 |
| Crossover probability | 0.8 |
| Mutation probability | 0.001 |

5.2. **Performance evaluation of emphGENESIS.** Now, the performance of *GENESIS* is evaluated by comparing with *AVIDO*. In this context, two uncertainty metrics are given consideration, namely, the presence and absence of (successful) transmission rate, respectively. These are regarded as subproblems in *GENESIS*

5.2.1. *Simulation Settings.* All nodes are set in simulator by considering that all have same transmission range(2cm). A random value between [0,0.8] is assigned to transmission rate between nodes. The simulation parameters are summarized in Table 6. For a certain n(number of total nodes), 10 instances are created for a certain number of n(Number of total nodes) and the result is the average value of 10 instances.

In Table 7, $TR_E$, $D_E$ represent successful transmission rate, CDS size in EVO-AUTO. On the other hand, $TR_G$, $D_G$ represent successful transmission rate, CDS size in *AVIDO*. It shows that the successful transmission rate decreases as the volume is increased, since the CDS size is increased. However, $TR_E$ ensures more successful transmission rate than $TR_G$. However, the size of CDS in $TR_E$ is greater than that of $TR_G$. Therefore, more resilience is achieved with $TR_E$ than $TR_G$. It is notable that $\frac{\sum_{i=1}^{T} \lambda_i \cdot TR_i^2}{D^2}$ are used to correlate two uncertainty metrics, namely transmission rate and minimum hop.

Table 7. Performance comparison of *GENESIS* with *AVIDO*)

| $m \times m$ | n | $TR_E$ | $TR_G$ | $D_E$ | $D_G$ |
|---|---|---|---|---|---|
| $50 \times 50$ | 60 | 0.52 | 0.60 | 22 | 30 |
| $60 \times 60$ | 100 | 0.45 | 0.55 | 35 | 41 |
| $70 \times 70$ | 180 | 0.40 | 0.47 | 53 | 59 |
| $80 \times 80$ | 250 | 0.35 | 0.45 | 68 | 77 |
| $100 \times 100$ | 400 | 0.33 | 0.41 | 72 | 86 |

5.3. Case Study and Application Scenario.

5.3.1. *Case Study.* The death toll from cancer is increasing due to the aging population, the increasing world population, and the increasing prevalence of cancer-causing behaviors such as smoking [21]. The World Health Organization (WHO) has concluded that cancer replaced heart disease as the overall leading cause of death in 2010[21]. The complete eradication of chronic diseases such as cancer[25] is still a long ways off. However, cancer can be treated in different ways depending on patient's medical condition and the type and location of the cancer. Common treatment methods are radiotherapy, chemotherapy, surgery, immunotherapy, and gene therapy[11][20][12]. Radiotherapy uses high energy radiation (X-ray, gamma ray or charged particles) to shrink or kill tumors. Chemotherapy delivers anticancer drugs to destroy or de-activate cancer cells. However, both therapies can be harmful to normal tissues as well and patients can suffer from serious health damage such as fatigue, hair loss, difficulty in eating, skin irritation, and vomiting. In hyperthermia, external or internal heating is provided to shrink tumors by destroying cancer cells or depriving them of their required substances. Combined trials[11]-[12] demonstrated that hyperthermia increases the sensitivity of cancer cells to chemotherapy or radiotherapy, indicating that it is a promising strategy for combination therapy for treating cancer. Hyperthermia enhances the performance of radiotherapy and chemotherapy, as long as the tissue temperature is maintained below a threshold value so as to avoid tissue damage due to excessive heat[11]-[12].

Body sensor networks (BSN)[26] have the potential to change the medical diagnosis system. One example is in vivo BSNs,[6] where sensors implanted inside the human body communicate wirelessly with a gateway for monitoring or diagnosis information. Because of its non-invasiveness, more accurate information, and utility in long-term monitoring, clinicians have accepted its usage in critical and sensitive health-care applications such as endoscopy capsules, deep-brain stimulation, cardiac pacemakers, artificial retinas, and core-body temperature calculators[19]. In vivo temperature sensors monitor the body's core temperature and communicate with a handheld device in real time[19][5][10]. Since NASA Goddard Space Flight Center and John Hopkins University developed the technology to calculate the body temperature of astronauts and relay the information to earth, it has received an immense amount of interest from the medicine science community. In vivo temperature sensors are widely used to calculate the body temperature of fire-fighters, deep sea saturation drivers, distance runners, and soldiers, and even American football teams have used the technology [19][5][10]. It is in this context that we were motivated to explore the use of in vivo-BSN in temperature scheduling as part of a combined therapy for treating cancer.

However, *in vivo* temperature sensors themselves generate heat in surrounding biological tissues by electromagnetic field induction and power dissipation. Therefore, the generated heat is compared to the known human electromagnetic wave safety level as defined by the IEEE

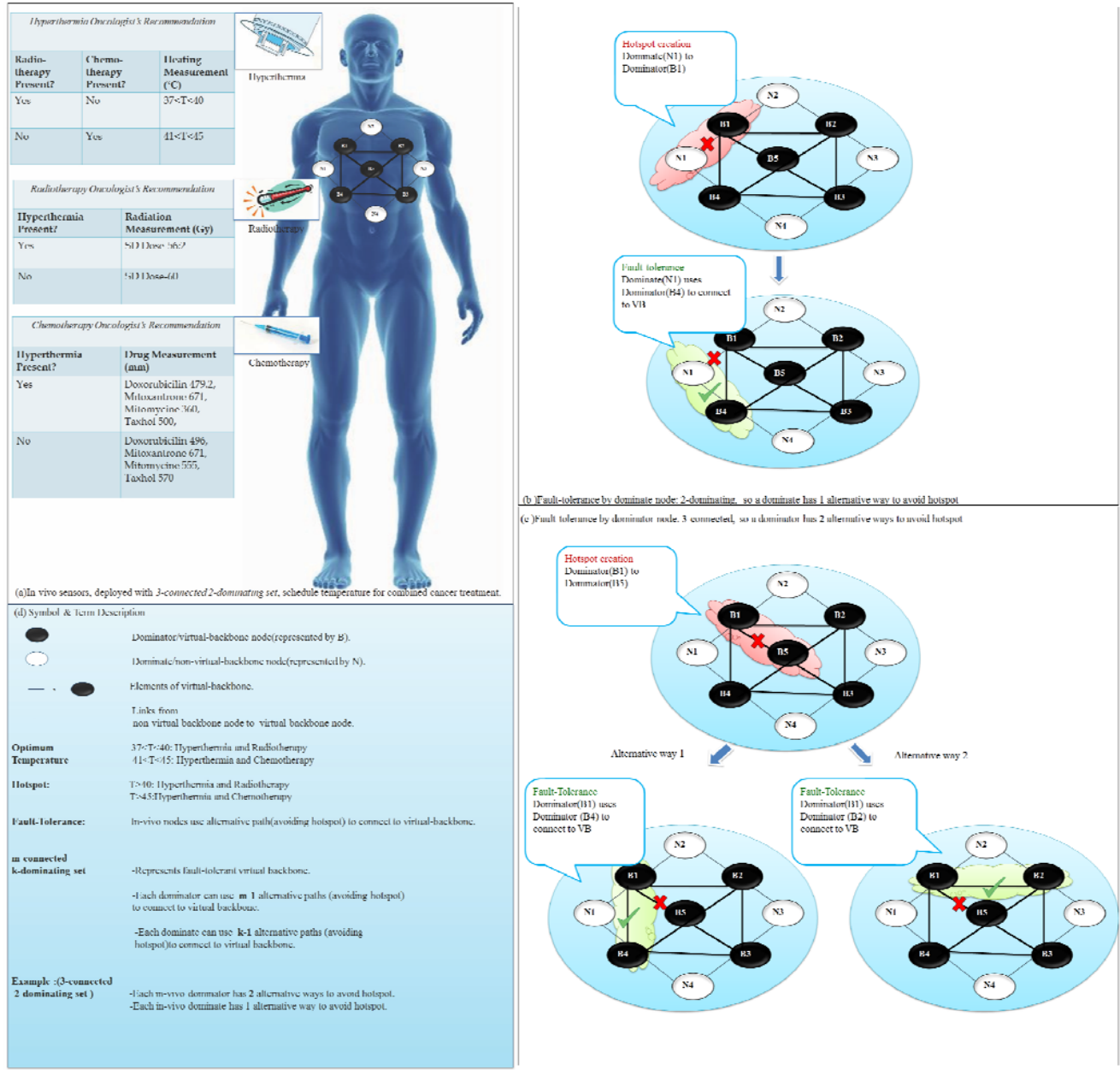

Figure 3. Case study and application scenario of Evolvable Autonomic Management in Combined Cancer Treatment with In-vivo Body Sensor Networks

Standard[1]. In this context, research on thermal awareness is given significant attention in in vivo research. In accordance with safety guidelines, thermal aware routing algorithms, namely TARA[23],LTR[3], LTRT[22],HPR[4] are proposed to schedule temperature via in vivo communication. However, the limitations of these algorithms motivate us for autonomic management approach.

5.3.2. *Application Scenario.* In this context, evolvable autonomic management becomes significant for uncertain thermal aware communication in an in-vivo BSN for two reasons. Firstly, it is very easy to deploy any other thermal routing protocols through it. Secondly, it is resilient to hotspot creation and uncertain communication in in-vivo BSN.

Clinical trials from several prominent combined trials conclude that hyperthermia enhances the performance of radiotherapy, if the temperature is maintained between 37°C and 40°C. Moreover, Hyperthermia is best suited for use with chemotherapy at temperatures between 41°C and 45°C. However, if the temperature exceeds the threshold, heating might cause serious cell damage, bacterial effect, and toxicity, as human cells become very sensitive at high temperatures[11]-[7]. An exemplary situation is considered, where hyperthermia, radiotherapy, and chemotherapy are being applied to a patient (Fig. 3). The recommendations of the hyperthermia, radiotherapy and chemotherapy oncologists are summarized in Fig. 3(a). In vivo sensors are deployed in the patient's body to schedule temperature changes by 3-connected 2-dominating set with minimum temperature. Symbol and term descriptions for this example formulation are given in Fig. 3(d).The heating used in hyperthermia might not be harmful to biological cells, if the temperature is below 40°C and 45°C for radiotherapy and chemotherapy, respectively. However, if hyperthermia exceeds the threshold, a hotspot might be created. However, the nearby in vivo sensors are able to sense the hotspot. Moreover, an in vivo sensor might create a hotspot itself after continual operation for an extended period of time. In this context, 3-connected 2-dominating sets help to avoid hotspots in the following ways:In Fig. 1(b), the VB (B1,B2,B3,B4,B5) is 2-dominating, so, if the link from dominate N1 to dominator B1 is in a hotspot, N1 might use alternative dominator B4 to connect to the VB.In Fig. 3(c), the VB (B1,B2,B3,B4,B5) is 3-connected, so, if a link from dominator B1 to dominator B5 is in a hotspot, B1 might use two alternative dominators (B4 or B2) to connect to the VB.

## 6. Conclusion

In this paper, an evolvable autonomic management technique is proposed for the resilience to uncertainty. In this context, at first, autonomic management problem is devised which is aimed at constructing $m$-connected $k$-dominating set with multiple uncertainty constraints. However, the absence of PTAS for original problem is justified its amenability to multi-objective evolutionary algorithm. Hence, a multi-objective evolutionary approach *GENESIS* is proposed, which facilitates resilient to uncertainties by evolving solutions of correlated uncertainty constraints, preceded by initial population from a greedy approach *AVIDO*. Extensive simulation is undergone to justify the supremacy of proposed greedy and evolutionary approach over the state-of-the art. Finally, an extensive case study is performed to implicate devised evolvable autonomic management approach on combined cancer treatment with the advent of in-vivo sensor networks.

## 7. Acknowledgment

Acknowledgments of people, grants, funds, etc. should be placed in a separate section before the reference list. The names of funding organizations should be written in full.

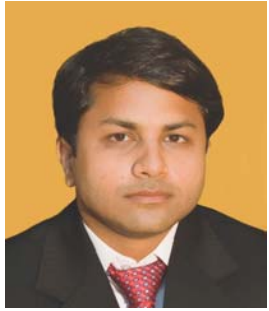

R. Kamal - graduated from Department of Computer Engineering of University of Dhaka in 2009. He has finished his Ph. D coursework on Department of Computer Engineering from Kyung Hee University, South Korea. He has worked as reviewer in IEEE WCNC, IEEE ICC, IEEE Globecom, IEEE Communications Magazine,IEEE Transactions on Image Processing